\documentclass{webofc}
\usepackage[varg]{txfonts}   
\begin{document}
\title{Luminosity determination for the proton-deuteron reaction using $pd\rightarrow$ $^{3}\hspace{-0.03cm}\mbox{He} \eta$ channel with WASA-at-COSY detector}
%
%

\author{
      \firstname{Magdalena} \lastname{Skurzok}\inst{1}\fnsep\thanks{\email{magdalena.skurzok@uj.edu.pl}} 
      \and
\firstname{Oleksandr} \lastname{Rundel}\inst{1}\fnsep\thanks{\email{o.rundel@doctoral.uj.edu.pl}} 
    \and
        \firstname{Aleksander} \lastname{Khreptak}\inst{1}\fnsep\thanks{\email{aleksander.khreptak@doctoral.uj.edu.pl}} \and
             \firstname{Pawe\l} \lastname{Moskal}\inst{1}\fnsep\thanks{\email{p.moskal@uj.edu.pl}} 
             }

\institute{Institute of Physics, Jagiellonian University, prof. Stanis{\l}awa {\L}ojasiewicza~11, 30-348 Krak\'{o}w, Poland
          }

\abstract{%
In this report one of the methods used for the luminosity determination for the experiment performed by WASA Collaboration to search for $^{3}\hspace{-0.03cm}\mbox{He}$-$\eta$ mesic nuclei is presented. The method is based on the analysis of $pd\rightarrow$ $^{3}\hspace{-0.03cm}\mbox{He} \eta$ process.}


%
\maketitle
\section{Introduction}
\label{intro}

\noindent Although strongly bound $\eta$- and $\eta'$-mesic nuclei, being one of the hottest topics in nuclear and hadron physics, have been postulated many years ago~\cite{HaiderLiu1}, until now their existence has not been experimentally confirmed.~Signals measured so far might be only interpreted as an indications of the $\eta$- and $\eta'$-mesic bound states~\cite{Sokol_2001,Budzanowski,Berger,Mayer,Smyrski1,Smyrski2,Smyrski3,MoskalSmyrski,Mersmann,Papenbrock}. Recently, several theoretical~\cite{Ikeno_EPJ2017,Xie2017,Fix2017,Barnea2017,Gal2017,Gal2015,Friedman,Kelkar_2016_new,Kelkar,Kelkar_new,Wilkin_2016,BassThomas1,BassThomas,Hirenzaki1,Nagahiro_2013,WycechKrzemien,Niskanen_2015,NagahiroTakizawa2006,NagahiroHirenzaki2005,Tshushima1,Tshushima2,Garcia} and experimental studies~\cite{Tanaka,Skurzok_NPA,Adlarson_2013,Acta_2016,Machner_2015,Krusche_Wilkin,Metag2017,Moskal_Acta2016,Moskal_AIP2017,Moskal_FewBody} are ongoing.

One of the most recent and most promising experiments dedicated to search for $\eta$-mesic Helium nuclei has been performed by the WASA-at-COSY Collaboration at the Forschungszentrum J\"ulich.~The measurements were carried out with high statistics and high acceptance with the WASA detection setup in deuteron-deuteron ($^{4}\hspace{-0.03cm}\mbox{He}$-$\eta$)~\cite{Skurzok_NPA,Adlarson_2013} and
proton-deuteron ($^{3}\hspace{-0.03cm}\mbox{He}$-$\eta$)~\cite{Proposal_2014} fusion reactions.~During experiments, the beam momentum was changed slowly and continuously around the $\eta$ production threshold in each of the COSY acceleration cycles.~This allowed us to search for the bound states via the measurement of the excitation functions for chosen decay channels. The analysis dedicated to search for $^{4}\hspace{-0.03cm}\mbox{He}$-$\eta$ mesic nuclei in $dd\rightarrow$ $^{3}\hspace{-0.03cm}\mbox{He} n \pi{}^{0}$ and $dd\rightarrow$ $^{3}\hspace{-0.03cm}\mbox{He} p \pi^{-}$ processes results in the upper limits of the total cross section at 90\% confidence level equal to about 3~nb and 6~nb, respectively~\cite{Skurzok_NPA}. 

The search for $^{3}\hspace{-0.03cm}\mbox{He}$-$\eta$ bound states was performed in May 2014 using a ramped proton beam with momentum changing continuously in the range of $1.426$-$1.635$~GeV and pellet deuterium target.~The range of the beam momentum corresponds to the range of excess energy $Q_{^3He\eta}$ from $-70$ to $+30~MeV$. The $^{3}\hspace{-0.03cm}\mbox{He}$-$\eta$ bound states~\cite{Acta_2016,Proposal_2014} are searched for via studies of the excitation functions for processes corresponding to the three mechanisms: (i) absorption of the $\eta$ meson by one of the nucleons, which subsequently decays into $N^{*}$-$\pi$ pair e.g.: $pd \rightarrow$ ($^{3}\hspace{-0.03cm}\mbox{He}$-$\eta)_{bound} \rightarrow$ $p p p \pi{}^{-}$, (ii) decay of the $\eta$ -meson while it is still "orbiting" around a nucleus e.g.: $pd \rightarrow$ ($^{3}\hspace{-0.03cm}\mbox{He}$-$\eta)_{bound} \rightarrow$ $^{3}\hspace{-0.03cm}\mbox{He} 6\gamma$ reactions and (iii) $\eta$ meson absorption by few nucleons e.g.: $pd \rightarrow$ ($^{3}\hspace{-0.03cm}\mbox{He}$-$\eta)_{bound} \rightarrow$ $ppn$. Almost two weeks of measurement brought the world's largest data sample for $^{3}\hspace{-0.03cm}\mbox{He}$-$\eta$. The total integrated luminosity and also its dependence on the excess energy above the $\eta$ production threshold was estimated based on the $pd \rightarrow$ $^{3}\hspace{-0.03cm}\mbox{He}\eta$ process. In this contribution we present the procedure of the luminosity calculation.

\section{Luminosity determination based on $pd\rightarrow$ $^{3}\hspace{-0.03cm}\mbox{He} \eta$ with WASA-at-COSY detector}
\label{sec-1}

The luminosity determination in the excess energy range is crucial for the normalization of the experimental excitation functions.~The luminosity is obtained from the analysis of the reaction that has a known cross section and allows one to calculate cross sections for other measured processes after obtaining the amount of events.

The experimental luminosity can be determined in the whole beam momentum range based on 
$pd\rightarrow ppn_{spec}$ reaction and, in addition, it can be also determined based on the
$pd\rightarrow$ $^{3}\hspace{-0.03cm}\mbox{He}\eta$ reaction in the range above the $\eta$ production threshold ($Q_{^3He\eta}>0$).
The whole $Q_{^3He\eta}$ range from $-70$ to $+30~MeV$ was divided into 40 bins, each one with the width of $2.5~MeV$.

Luminosity estimation from the $pd\rightarrow$ $^{3}\hspace{-0.03cm}\mbox{He}\eta$ reaction is the first stage of the data analysis.~Events corresponding to this reaction are identified by $^{3}\hspace{-0.03cm}\mbox{He}$ tracks registered in the forward part of the WASA detector. $^{3}\hspace{-0.03cm}\mbox{He}$ identification is carried out with the $\Delta E$-$E$ method and kinetic energy reconstruction from the energy deposited in the detector.~The angles are reconstructed based on registered tracks with the algorithms implemented by the WASA-at-COSY team and used in the previous analysis.
%
%
%
%
%
%
%
%
%
%

Kinetic energy $^{3}\hspace{-0.03cm}\mbox{He}$ is reconstructed based on the value of energy deposited in $FRH1$ layer of the forward detector. For this layer energy deposition by $^{3}\hspace{-0.03cm}\mbox{He}$ ions is the largest. The kinetic energy is determined from a fit of $\theta(E_{kin})$ spectrum obtained from Monte Carlo simulation with a formula:

\begin{equation}
E_{kin} = f_0(\theta) + f_1(\theta)E_{FRH1},
\end{equation}

where $f_0$ and $f_1$ are squared functions of the scattering angle $\theta$:

\begin{equation}
f_i(\theta) = a_i + b_i \theta + c_i \theta^2.
\end{equation}

\begin{figure}[h!]
\centerline{
\includegraphics[width=6.5cm,height=5.0cm]{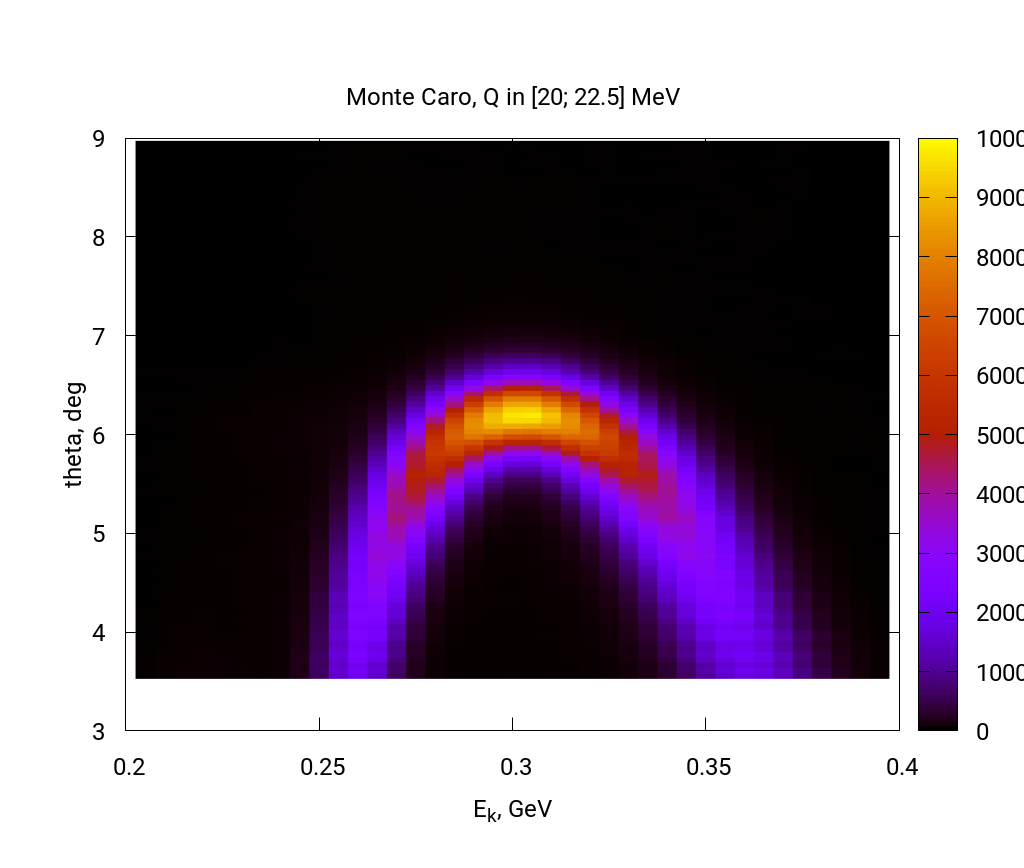}
\includegraphics[width=6.5cm,height=5.0cm]{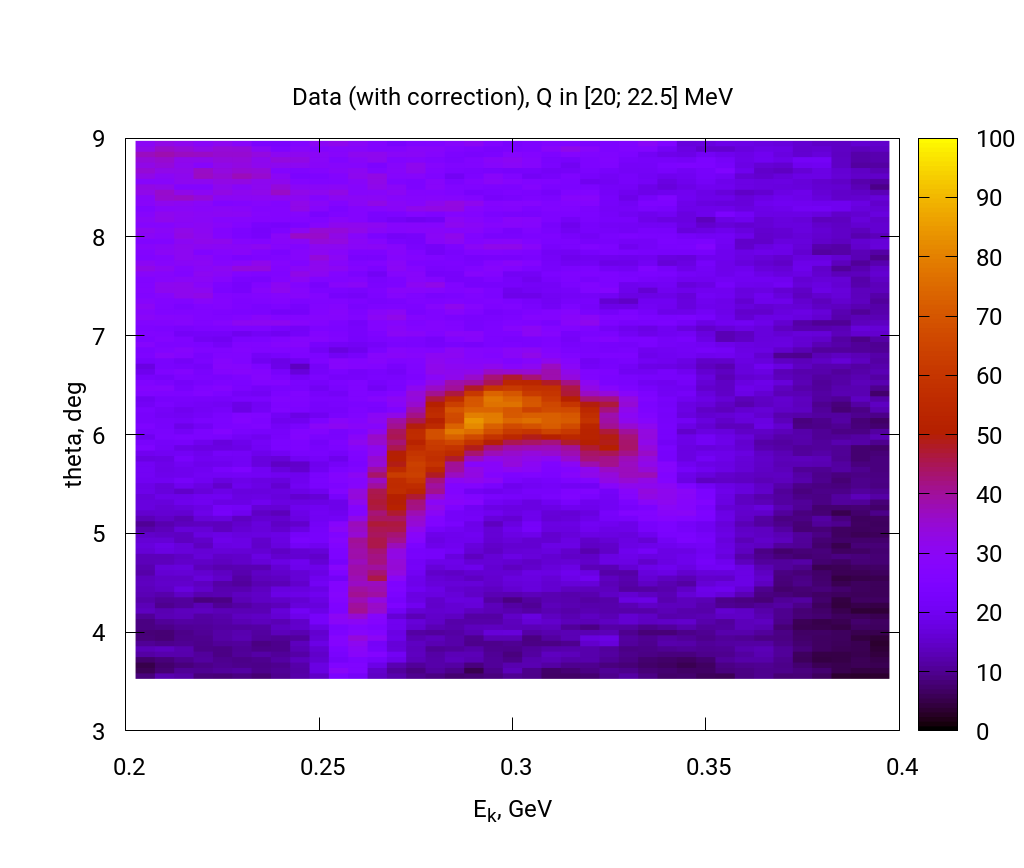}
}

\caption{The 2D distribution of $^{3}\hspace{-0.03cm}\mbox{He}$ energy (horizontal) and $\theta$ angle (vertical) for the excess energy interval
\mbox{$Q_{^3He\eta}\in(17.5,20)MeV$}. Left side: reconstruction algorithm run over Monte Carlo simulation. Right side: result of data analysis performed with assuming the correction constants $off_{P}=4.6~MeV/c$ and $off_{E}=2~MeV$.}
\label{Fig:kin-2d}
\end{figure}

The fit contains six free parameters. For data, the beam momentum is obtained as a function on accelerator's time-in-cycle that precisely describes the momentum changes but contains unknown offset. It was also assumed that the kinetic energy for data may differ from the reconstructed value by a small correction constant. These two constants were obtained from kinematic distributions (Fig.~\ref{Fig:kin-2d}) comparing positions of $E-\theta$ distributions for simulation and data. The values obtained this way are equal to  $off_{P}=4.6~MeV/c$ and $off_{E}=2~MeV$.

\begin{figure}[h!]
\centerline{
\includegraphics[width=6.5cm,height=5.0cm]{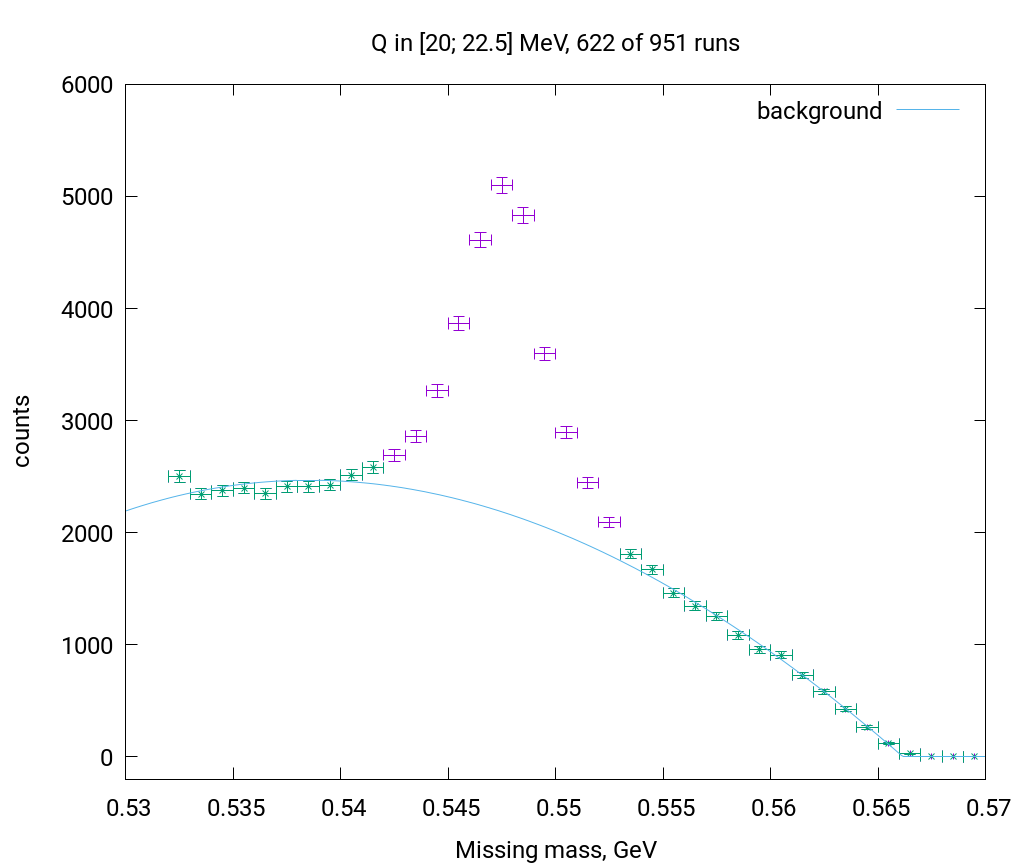}
\includegraphics[width=6.5cm,height=5.0cm]{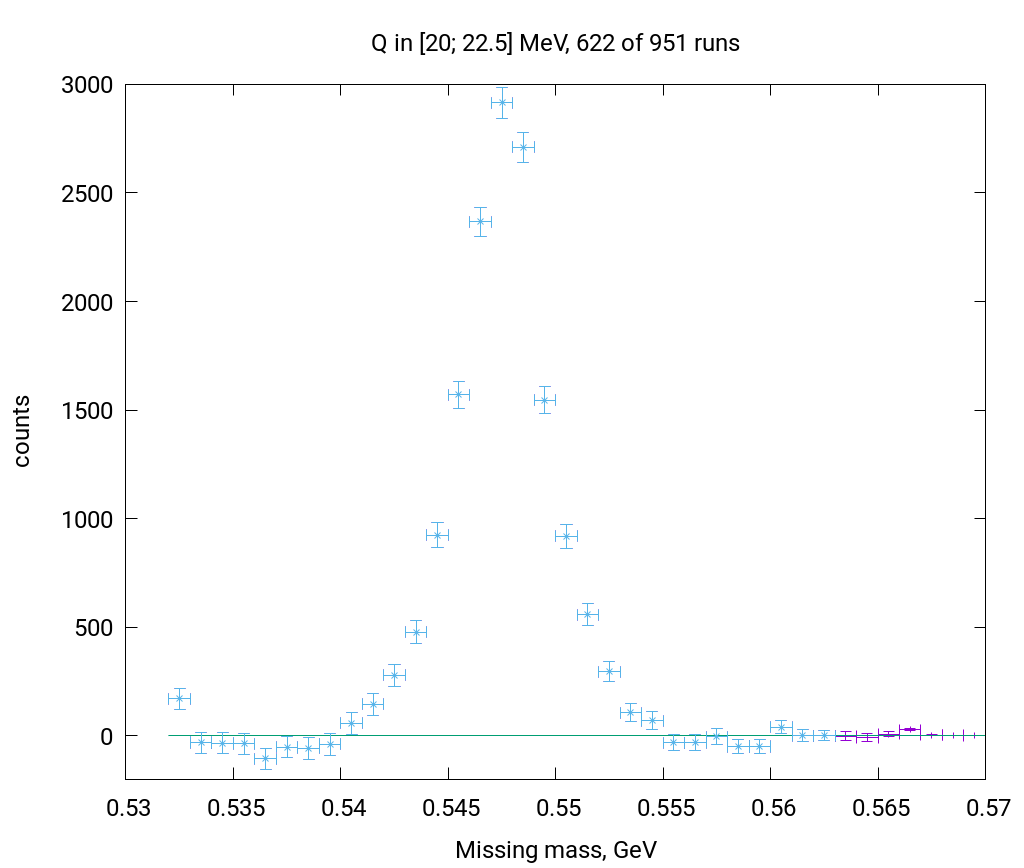}
}
\caption{The $^{3}\hspace{-0.03cm}\mbox{He}$ missing mass spectrum for the excess energy interval
\mbox{$Q_{^3He\eta}\in(20,22.5)MeV$}. Left side: the background around the $\eta$ creation peak is
fit with a polynomial. Right side: missing mass after the background subtraction (for obtaining
the amount of $\eta$ creation events).}
\label{Fig:spectrum}
\end{figure}

\begin{figure}[h!]
\centerline{
\includegraphics[width=6.5cm,height=5.0cm]{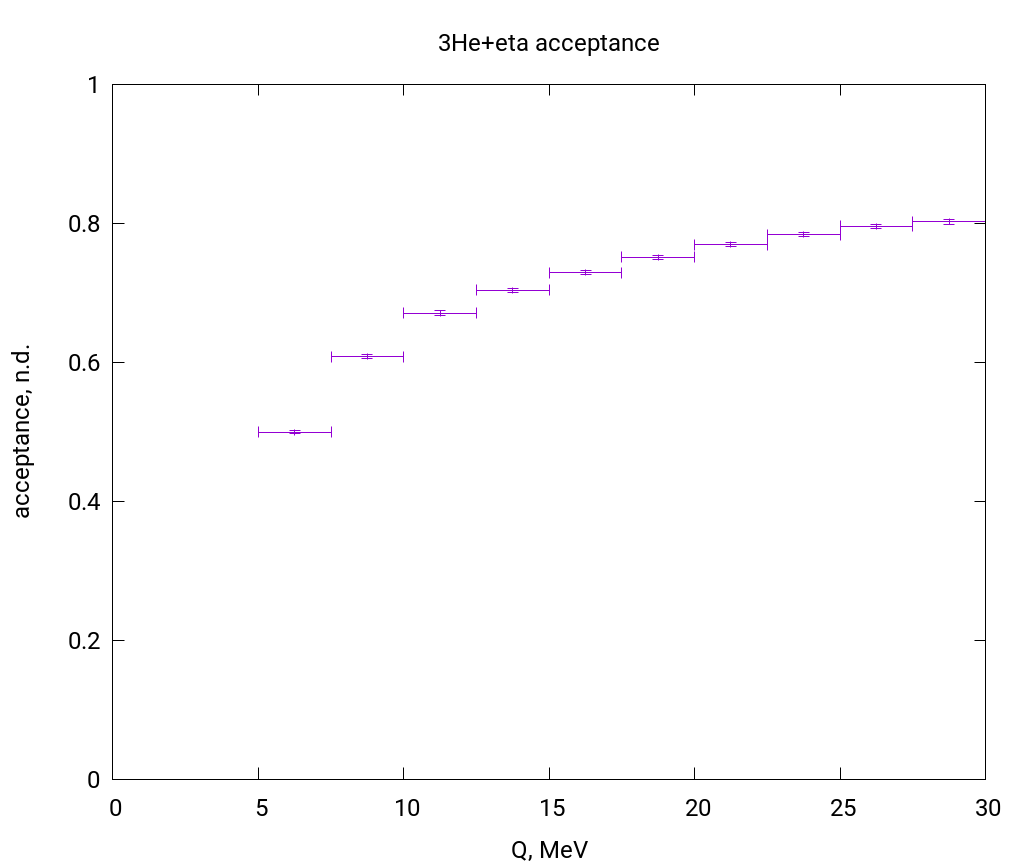}
}
\caption{The acceptance values for $pd\rightarrow^{3}\hspace{-0.03cm}\mbox{He}\eta$ reaction obtained from Monte Carlo simulations.}
\label{Fig:scceptance}
\end{figure}

\indent For obtaining the amount of events corresponding to $\eta$ creation, the $^{3}\hspace{-0.03cm}\mbox{He}$ missing mass spectra have been analyzed for each excess energy $Q_{^3He\eta}$ intervals.~These events are visible as a peak around the value equal to the mass of the $\eta$ meson.~The value of integrated luminosity for each $Q_{^3He\eta}$ bin was obtained from the formula $L=\frac{N}{\epsilon \sigma}$, where the amount of events $N$ is extracted 
from the $\eta$ creation peak area (Fig.~\ref{Fig:spectrum}), the acceptance $\epsilon$ is determined based on Monte Carlo simulation (Fig.~\ref{Fig:scceptance}) and the total cross section $\sigma$ is taken from other experiments \cite{Mersmann}.


\begin{figure}[h!]
\centerline{
\includegraphics[width=8.5cm]{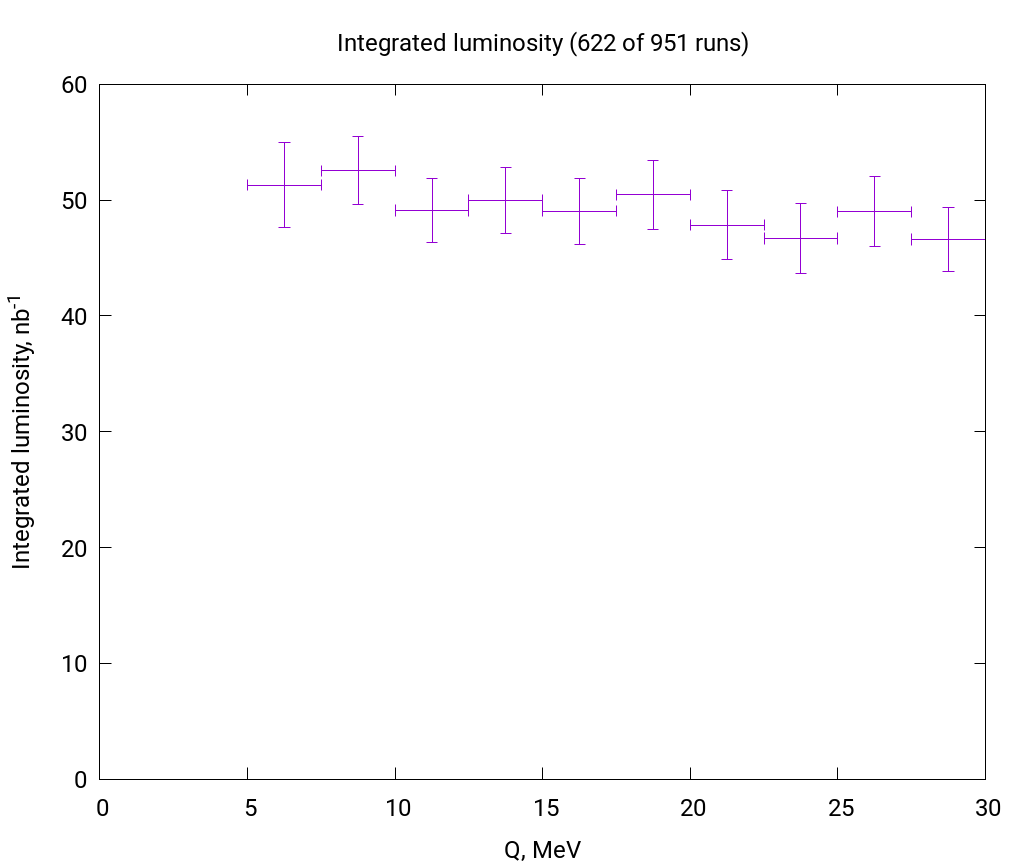}}
\caption{Integrated luminosity determined based on $pd\rightarrow$ $^{3}\hspace{-0.03cm}\mbox{He}\eta$ reaction for the excess energy range of \mbox{$Q_{^3He\eta}>0$}. The luminosity was calculated for $65~\%$ of collected data.}
\label{Fig:luminosity}
\end{figure}

The obtained integrated luminosity for each $Q_{^3He\eta}$ bin is shown in Fig.~\ref{Fig:luminosity}.
Total integrated luminosity of about $3~pb^{-1}$ was estimated for whole excess energy range assuming that
excess energy intervals for $Q_{^3He\eta}<0$ have the same values as determined for $Q_{^3He\eta}>0$ (about $50~nb^{-1}$) and taking into account the amount of data that has not been analyzed yet.
If we estimate the time of measurement as $10^6~s$ then obtained value of average luminosity becomes $3~10^{30}~cm^{-2}s^{-1}$.
This value is in agreement with value given in the proposal~\cite{Proposal_2014}.

\section{Summary}
One of two planned luminosity determination analyses for the experiment performed with WASA-at-COSY to search the $^{3}\hspace{-0.03cm}\mbox{He}$-$\eta$ bound states in proton-deuteron fusion has been carried out.
The analysis was performed for the $pd\rightarrow^{3}\hspace{-0.03cm}\mbox{He}\eta$ reaction.
The estimated value of average luminosity is equal to  $3~10^{30}~cm^{-2}s^{-1}$ and is in agreement with value presented in proposal~\cite{Proposal_2014}.
The obtained luminosity will be used for the normalization of the excitation functions for processes in which the $^{3}\hspace{-0.03cm}\mbox{He}$-$\eta$ nuclei are searched for. The data analysis is in progress.
The estimated integral luminosity value for this experiment is about $3~pb^{-1}$ and is the best statistics ever gathered for such experimental conditions.
%
%
%
%
%
%
\section{Acknowledgements}

\noindent We acknowledge the support from the Polish National Science Center through grant No. 2016/23/B/ST2/00784 and Jagiellonian University through grant DSC 2011 for financing young scientist and PhD students No. 142/F/OR/2016.
%
%
%
%
%
%

%
%
%

\end{document}